\documentclass{emulateapj}

\shorttitle{Luminosity-Luminosity Correlations}
\shortauthors{Singal et al.}

\def\beq{\begin{equation}}
\def\eeq{\end{equation}}

\def\d{{\delta}}
\def\g{{\gamma}}

\def\x,y{\{x,y\}}

\begin{document}

\title{LUMINOSITY-LUMINOSITY CORRELATIONS IN FLUX-LIMITED MULTIWAVELENGTH DATA}

\author{J. Singal\altaffilmark{1}, V. Petrosian\altaffilmark{2}$^,$\altaffilmark{3}, J. Haider\altaffilmark{1}, S. Malik\altaffilmark{1}}

\altaffiltext{1}{Physics Department, University of Richmond\\138 UR Drive, Richmond, VA 23173}
\altaffiltext{2}{Department of Physics and Kavli Institute for Particle Astrophysics and Cosmology, Stanford University\\382 Via Pueblo Mall, Stanford, CA 94305-4060}
\altaffiltext{3}{Also Department of Applied Physics}

\email{jsingal@richmond.edu}

\begin{abstract}
We explore the general question of correlations among different waveband luminosities in a flux-limited multiband observational data set.  Such correlations, often observed for astronomical sources, may either be intrinsic or induced by the redshift evolution of the luminosities and the data truncation due to the flux limits.  We first address this question analytically.  We then use simulated flux-limited data with three different known intrinsic luminosity correlations, and prescribed luminosity functions and evolution similar to the ones expected for quasars.  We explore how the intrinsic nature of luminosity correlations can be deduced, including exploring the efficacy of partial correlation analysis with redshift binning in determining whether luminosity correlations are intrinsic and finding the form of the intrinsic correlation.  By applying methods that we have developed in recent works, we show that we can recover the true cosmological evolution of the luminosity functions and the intrinsic correlations between the luminosities.  Finally, we demonstrate the methods for determining intrinsic luminosity correlations on actual observed samples of quasars with mid-infrared, radio, and optical fluxes and redshifts, finding that the luminosity-luminosity correlation is significantly stronger between mid-infrared and optical than that between radio and optical luminosities,  supporting the canonical jet-launching and heating model of active galaxies. 
\end{abstract}

\section{Introduction} \label{intro}

\begin{figure}
\includegraphics[width=3.5in]{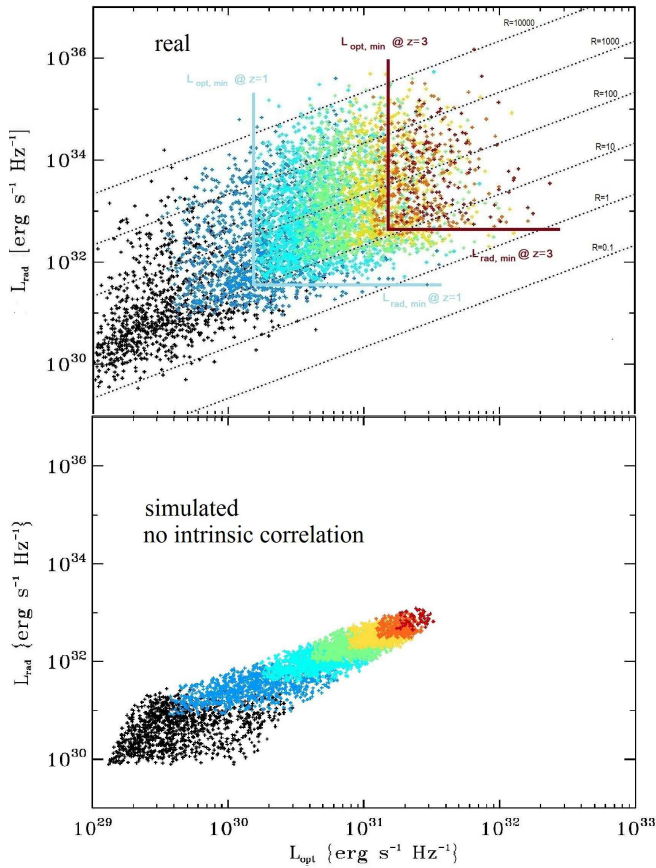}
\caption{ Two examples of observed optical-radio luminosity scatter diagrams from flux-limited data sets: ({\bf top}) quasars from \citet{QP2}, and  ({\bf bottom}) a simulated data set subject to similar flux limits and known input parameters, but with {\it no intrinsic luminosity-luminosity correlation} developed in \S \ref{sims}.  Colors represent different redshift bins.  Black points are $z \leq 0.5$, dark blue points are $0.5<z \leq 1.0$, light blue points are $1.0<z \leq 1.5$, green points are $1.5<z \leq 2.0$, yellow points are $2.0<z \leq 2.5$, orange points are $2.5<z \leq 3.0$, and red points are $z > 3.0$.  Also shown for the real data are lines of constant  ratio of the 5 GHz radio luminosity to the 2500 \AA \, optical luminosity, and the limiting luminosities for inclusion in the sample at example redshifts of $z=1$ and $z=3$.  It is clear that selection and redshift evolutions can induce a correlation between the different waveband luminosities that is not intrinsic.  Some of the sharpness of the boundaries observed in the simulated data is due to a high luminosity cutoff included in the simulated data to make the number of required data points manageable. }
\label{firstfig}
\end{figure}

When dealing with multiwavelength observations of astrophysical sources the question often arises whether the emissions in different wavebands (e.g. optical, radio, infrared, X-ray, gamma-ray, etc.) are  correlated.  Determining the intrinsic correlations between these emissions is crucial for addressing large variety of scientific questions, e.g.~the relation between the emission processes and the sites and mechanisms of the acceleration of particles (or more generally the energizing of the plasma) responsible for these emissions.  A common practice is to plot luminosities in two bands against each other for a sample of observed sources and determine the luminosity-luminosity (hereafter $L$--$L$) correlation empirically.  However, more often than not such samples  include sources with a large range of distances such as extragalactic sources with a range of redshifts.\footnote{Exceptions arise in dealing with clusters of sources with sizes much smaller than their distance  (e.g.~Galactic star clusters, sources in distant individual galaxies or clusters of galaxies).}  Such samples are always subject to observational selection effects that truncate the data.  The most common truncation arises in flux-limited data, where the fact that lower (higher) luminosities in both bands are dominated by sources at lower (higher) redshifts introduces a significant artificial correlation in the observed luminosities \citep[e.g.][]{Pavlidou12,Ski,FB83,Khembavi86,Chanan83}.  The situation is even more complicated, however, with extragalactic sources, where in addition to the observational selection effects, the different luminosities may undergo similar or different cosmological luminosity evolution which can induce additional $L$--$L$ correlation \citep{CR1}.  Figure \ref{firstfig} shows two examples of $L$--$L$ scatter diagrams obtained from flux-limited observed data (top panel) and simulated data described below (bottom panel).  The top panel is from an actual observed data while the bottom panel is from a simulated observed data set described below.  In the latter case the population has no intrinsic $L$--$L$ correlation by design yet displays a strong observed $L$--$L$ correlation.  \citet{CR1}, using partial correlation coefficients and Efron-Petrosian non-parametric methods \citep{EP92, EP99}, showed that most but not all of the observed correlation in the top panel is induced by the selection process.

In this work we explore the question of to what extent observed correlations in multiwavelength flux-limited data are indicative (or not) of intrinsic correlations, and develop and verify techniques for directly determining correlations and distributions.  In \S \ref{analytic} we show analytically the extent to which i) truncations due to flux limits of the samples and/or ii) luminosity evolutions induce artificial $L$-$L$ correlation and the dependence of these effects on the characteristics of the luminosity functions (LFs).  

It should be noted that the questions under considerations here are relevant not only for $L$--$L$ correlations but are important for exploring the correlation, or generally the relation, between any two characteristics (or variables) both of which depend on and are obtained from the values of a third independent characteristic.  In such a case partial correlation coefficients (based on, for example, Pearson or Kendall statistics) must be used as explored here.  In astrophysical sources this applies to all extensive characteristics such as luminosity, mass or size, whose values can only be obtained with the measurement of their distances, which are subject to data truncation and in the case of extragalactic sources are affected by cosmological evolutions mentioned above. Thus the procedures and results described here for $L$--$L$ correlation is relevant for considerations of correlations between any two (similar or different) pairs of extensive characteristics.

In the next section we present some analytic calculations showing the degree by which different effects mentioned above induce artificial $L$--$L$ correlation.  In \S \ref{sims} we introduce and explore simulated data sets with known intrinsic characteristics of the LF with different degrees of intrinsic correlation between different waveband luminosities. In \S \ref{partcor} we explore the efficacy of partial correlation analysis with redshift binning in determining whether luminosity correlations are intrinsic.  In \S \ref{ep} we demonstrate that techniques applied in recent works \citep{QP1,BP1,BP2,QP2,BP3,QP3}, based on extensions of methods first proposed by Efron and Petrosian \citep{EP92,EP99} can recover the intrinsic distributions and correlations of the luminosities and redshifts in flux-limited multiwavelength data, and show that the intrinsic $L$--$L$ correlations can be deduced by considering the correlations between the de-evolved luminosities.  In \S \ref{real} we demonstrate the use of partial correlation analysis and the determination of the intrinsic $L$--$L$ correlations for two real multiwavelength data sets consisting of radio and optical and mid-infrared and optical observations.  We summarize the main results in \S \ref{disc}.

\section{Analytical Considerations} \label{analytic}

\begin{figure}
\includegraphics[width=3.5in]{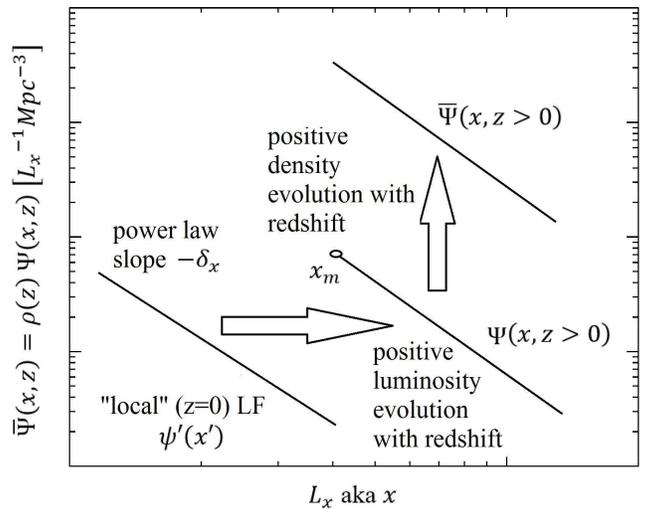}
\caption{A graphical sketch of some characteristics of a single-waveband differential LF ${\bar \Psi}\!(x,z)=\Psi\!(x,z)\rho\!(z)$ where $\rho\!(z)$ is the redshift density evolution and $\psi'_{x}\!(x')$ is the local LF, with $x'$ being the ``local luminosity'' -- that is the luminosity with its redshift evolution taken out.  Here $x$ is being used to represent the luminosity in the particular waveband, following the notation of \S \ref{analytic}, and all functions of the redshift $z$ could equivalently be expressed as functions of a cosmological distance measure $r$ as in that section.  $x_m$ represents the minimum luminosity of objects that could be present in the survey at some $z>0$ given the flux limit of the survey $f_{l,x}$.  The LF extends to luminosities below this value but is not directly probed by the survey in question.}
\label{LFfig}
\end{figure} 

Let us consider the general trivariate differential LF ${\bar \Psi}(x,y,r)$ where $x$ and $y$ stand for dimensionless (for algebraic convenience) luminosities ($\{x,y\}=L_{\{x,y\}}/L_0$), in two different photon energy bands (where the ${x,y}$ notation indicates that the equation in question applies to either $x$ or $y$), and $r$ stands for a measure of the distance of the object, which for extragalactic sources depends on redshift $z$ (or $Z \equiv1+z$).  $r$ can be the co-moving metric distance $D_C(z)$ or the luminosity distance $D_L(z)$.  In what follows we will use the last choice, i.e. $r$ will stand for $D_L$.\footnote{Note that in a static Euclidean case (for example if one is dealing with Galactic or nearby extragalactic sources) all these measures of distance are equivalent.}  

The ``differential'' LF quantifies the differential number of objects per infinitessimal bin in the relevant parameters, so obtaining the total number of objects in some range of parameter values involves integrating the differential LF over those ranges, while a ``cumulative'' LF would quantify the total number of objects above or below a certain value of a parameter, that is an integration over a certain range with a limit of 0 or $\infty$.  The luminosities in a sample are calculated from the observed fluxes $f_x, f_y$. We express these  dimensionless fluxes in units of fiducial flux $f_0=L_0/[4\pi(c/H_0)^2]$ so that we get $\{x,y\}=r^2f_{\{x,y\}}$.

Without loss of generality we can write, for the full differential LF
\beq\label{PsiRho}
{\bar \Psi}(x, y, r)=\Psi(x,y,r)\rho(r),
\eeq
with $\rho(r)$ describing the differential density evolution --- the change in number density of objects, and $\Psi(x,y,r)$ describing the dependence on the luminosities and redshift.  This simply splits the full LF of both luminosities into a portion quantifying the object density evolution and a portion quantifying everything else.  The differential density evolution $\rho(r)$ is related to the cumulative luminosity evolution $\sigma(z)$ by
\beq
{ {d\sigma(r)} \over {dr} }= \rho(r){dV\over dr}.
\label{ssigeq}
\eeq

If we then consider a sample of sources with flux limits $f_{l,x}$ and $f_{l,y}$ (in units of $f_0$), they would have minimum luminosities for inclusion in the sample as a function of $r$
\beq\label{limits}
 x_m(r)=r^2f_{l,x},\,\,\,\, 
y_m(r)=r^2f_{l,y}.
\eeq
Figure \ref{LFfig} graphically sketches some of the characteristics of a (single-waveband for clarity) LF when parameterized in this way.  The observed distribution of the sample --- that is the number density of objects in the parameter space of $x$, $y$, and $r$ --- is then related to these intrinsic luminosity distributions as
\beq
\label{ObsDist}
{{d^3N} \over {dx \, dy \, dr}}={ {d\sigma(r)} \over {dr} } \, \Psi(x, y, r) \, \Theta(x-x_m) \, \Theta(y-y_m)
\eeq
where $\Theta(b)$ is the step function (=1, for $b>0$ and =0 otherwise), and the distance related distribution (cumulative density evolution).  For convenience we also define bivariate observed luminosity distribution over just $x$ and $y$ luminosities (i.e. integrated over all $r$)
\beq\label{obsLDist2}
{{d^2N} \over {dx \, dy}} \equiv N(x,y)=\int_0^{r_0} (d^3N/dx \, dy \, dr) \, dr
\eeq
(where $r_0$ is the maximum distance accessed by the survey), and mono-variate distributions over just one luminosity
\beq\label{ObsLDist1}
{{dN} \over {{\{dx,dy\}}}}\equiv N(\{x,y\})=\int_{\{y_m,x_m\}}^\infty N(x,y) \, d\{y,x\}.
\eeq
We use the observed moments of these distributions to determine the correlation between the two luminosities. For example, the observed average value of luminosity $x$ at each fixed luminosity $y$ value is:
\beq\label{average1}
\langle x(y) \rangle={\int_{x_m}^\infty x \, d^2N/(dx \, dy) \, dx\over dN/dy}.
\eeq
(because dN/dy is the number (density) of objects at that $y$ value, thus the denominator of the average), and we can determine if the observed average value of $x$ depends on or is independent of this $y$ value, which is an indication of an observed correlation or lack thereof, respectively, between $x$ and $y$.

We start by assuming that the luminosities are uncorrelated (i.e. $x$ and $y$ are independent of each other), and that both are independent of $r$; i.e. there is no luminosity evolution, and see if an observed correlation is induced.  In this case we can separate the variables as ${\bar \Psi(x, y,r)}=\psi(x)\psi(y)\rho(r)$.  Then {\it if and only if} the data are not truncated $x_m=0$ and $r_0=\infty$ and equation \ref{average1} simplifies to:
\beq\label{intrn}
\langle x \rangle={\int_0^\infty x \, \psi(x) \, dx\over \int_0^\infty \, \psi(x) \, dx}=x_{\rm int}.
\eeq
 and $x_{\rm int}$ is a constant independent of $y$, thus resulting in no observed correlation between $x$ and $y$.  

In what follows we will add the effects of data truncation and consider what one would {\it observe} $\langle x(y) \rangle$ to be given the data at hand.  We will consider several cases starting with the (mathematically) simplest case.\begin{enumerate}

\item 

{\it Simple Power law LFs and No Luminosity Evolution:}

Here $\psi(x)=\phi_x x^{-\d_x}\Theta(x-x_0)$ (similarly $\psi(y)=\phi_yy^{-\d_y}\Theta(y-y_0)$). The no luminosity evolution implies that $\phi_x, x_0$ and $\d_x$ are independent of $r$.  The truncation due to flux limits introduces distances $r_{0,x}=\sqrt{x_0/f_{l,x}}$ and  $r_{0,y}=\sqrt{y_0/f_{l,y}}$ below which the sample is not truncated. We assume that  $r_{0,x}<r_{0,y}$.  The intrinsic average value (and the value observed for un-truncated data) is $x_{\rm int}=x_0(\d_x-1)/(\d_x-2)$ obtained by putting these LF forms into equation \ref{intrn}.  But for the truncated data the {\it observed} average values are, using these LFs and limits in evaluating equation \ref{average1}:
\beq\label{average2}
\langle x(y) \rangle={\int_0^{r_{0,x}} { {d\sigma} \over {dr} } dr \, \int_{x_0}^\infty dx \, x^{1-\d_x} +
\int_{r_{0,x}}^{\zeta} { {d\sigma} \over {dr} } dr \, \int_{x_m}^\infty dx \, x^{1-\d_x}\over
\int_0^{r_{0,x}} { {d\sigma} \over {dr} } dr \, \int_{x_0}^\infty dx \, x^{-\d_x} +
\int_{r_{0,x}}^{\zeta} { {d\sigma} \over {dr} } dr \, \int_{x_m}^\infty dx \, x^{-\d_x} }.
\eeq
where 
\beq\label{zzeta}
\zeta \equiv r_{0,y}\sqrt{y/y_0}>1.
\eeq
In order to evaluate these integrals we need the functional form of $d\sigma/dr$ which involves the product of two functions; the density evolution and the co-moving volume. In general this product is a complex function of the above measures of distance, in particular the $D_L$ being used here. Let us assume that we can approximate this with a power law, $d\sigma/dr\propto r^{2+\beta}$, with $\beta$ presenting roughly an evolution index. We then have, performing the integrations
\beq\label{average3}
\langle x(y) \rangle=x_{\rm int}\times{1+(3+\beta)\int_1^\zeta\eta^{2+\g}d\eta\over
1+(3+\beta)\int_1^\zeta\eta^\g d\eta}\,\,\,\, {\rm with}\,\,\,\, \g=4+\beta-2\d_x.
\eeq
so that the result depends primarily on the index $\g$. For $\g>-1$ ($\beta-2\d_x>-5$) the average value starts from near unity and rises quickly as $\langle x(y) \rangle\propto y$ with increasing $y$, while in the opposite limit of $\g<-3$ we get $\langle x(y) \rangle\sim$ const., and in between it varies more slowly than linearly with $y$. This indicates that in general data truncation induces some correlation between the luminosities and this correlation becomes stronger for larger values of the density evolution index $\beta$ and flatter LFs (smaller $\d_x$). This is as expected because both these effects result in a greater segregation of high and low luminosity sources at high and low redshifts, respectively, in the $L$--$L$ scatter diagrams as shown in Figure \ref{firstfig}. 

\item 

{\it  Broken Power Law LFs}

If broken power law applies only to one variable, say break of the $x$ LF at $x_{\rm br}$, then as evident from the above analysis the shape of the other LF (namely $y$) is unimportant, and the only complication is that in equation \ref{average2} we get three integrals in both the numerator and the denominator (the second integral gets divided into two at the break luminosity).  As indicated above a steeper LF induces weaker correlation, thus we expect that a steepening of the LF at higher luminosities, which is often the case for most astronomical sources, will reduce this effect. This can be seen by considering a very large steepening (i.e.~a large increase in value of $\d_x$ instead of changes of order unity seen in AGNs) which essentially sets a ceiling for the average near a value at the break luminosity  ($\langle x(y) \rangle\rightarrow x_{\rm br}$).

Now if the other LF also suffers a break (steepening at $y_{\rm br}$ as is common) then the integration limits become complicated depending on the relative values of the break luminosities and relative values of high luminosity indexes. In this case a numerical calculation, for specific parameters of the LFs, is required. 

\item

{\it Effects of Luminosity Evolution}

If the sources undergo luminosity evolution in one or both wavebands, we can express this luminosity dependence on redshift/distance with forms $x=x'g_x(r)$ and $y=y'g_y(r)$.  Here $x'$ and $y'$ will be referred to as the ``local luminosities,'' meaning that they are the luminosity values an object would have if the redshift evolution of the luminosity were taken out (provided we normalize the evolution function so that $g(0)=1$).  Now luminosity evolution in $x$ means that $\psi(x)$ is no longer independent of $r$, and evolution in $y$ necessitates that $\psi(y)$ is also not which means it does not come out of the $dr$ integrals and divide out when evaluating equation \ref{average1}.  In order to obtain variables independent of $r$ we can make the variable change in the integrals to $x'=x/g_x(r)$ and $y'=y/g_y(r)$.  We then get an equation very similar to equation \ref{average2} with $x$ in the integrals replaced by $x'$ and $d\sigma/dr$ changed to $d\sigma/dr\times g_x(r)$ (because $dx = dx' \times g_x(r)$).  Assuming positive luminosity evolution with redshift this increases the $\beta$ index, which as mentioned above increases the variation of the average observed $x$ with $y$ and thus the observed correlation of the (non-intrinsically correlated) luminosities.  A possible counter effect, however, is that with this variable change the $y$ value in equation \ref{zzeta} is reduced (assuming positive luminosity evolution in $y$) to $y'$ which generally acts toward reducing the observed correlation.  Since the designations of which band is $x$ and which is $y$ for this analysis is arbitrary, the effects must be symmetrical to this distinction.  Thus, perhaps counterintuitively, the induced observed correlation due to luminosity evolution is highest when the luminosity evolutions in the respective bands are the most divergent.   

\end{enumerate}

The above results show that for flux-limited, multi-waveband data sets, observational selection effects induce an artificial correlation between luminosities, the exact degree of which depends on the particular functional forms of the LFs and the two luminosity evolutions, with greater difference in the latter inducing greater artificial correlations between the luminosities.  The simulations discussed below show that these artificial $L$--$L$ correlations are indeed induced for such data sets.

\section{Simulated Data Sets} \label{sims}

\begin{figure}
\includegraphics[width=3.5in]{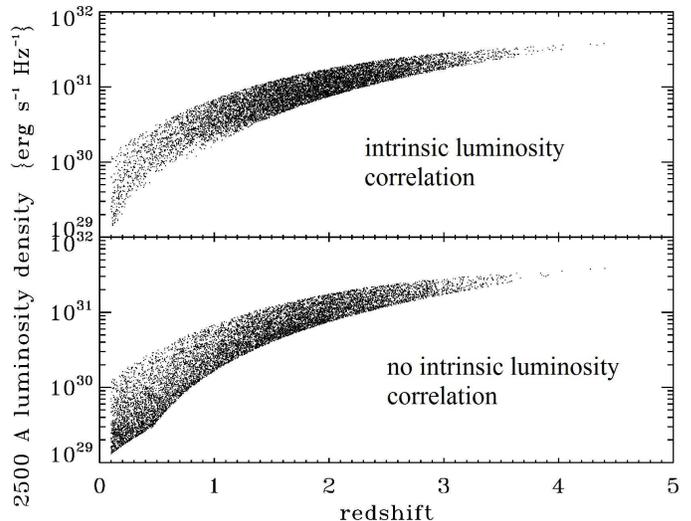}
\caption{The $i$ band optical vs. 1.4 GHz radio fluxes for the observed simulated data sets, for the case of intrinsically correlated ({\bf top} --- $\alpha=1.0$) and intrinsically uncorrelated ({\bf bottom} --- $\alpha=0$) radio and optical luminosities.  For clarity and ease of presentation here we show fluxes of 10,000 randomly selected sources. {\bf A possibly apparent discontinuity in radio flux at low optical fluxes in the uncorrelated case is a visual artifact of the visually compressed $y$-axis interfacing with the actual distribution of fluxes of the observed sources and the selection of 10,000 objects for visual purposes and does not represent an actual discontinuity in the data. } }
\label{simfluxflux}
\end{figure} 

\begin{figure}
\includegraphics[width=3.5in]{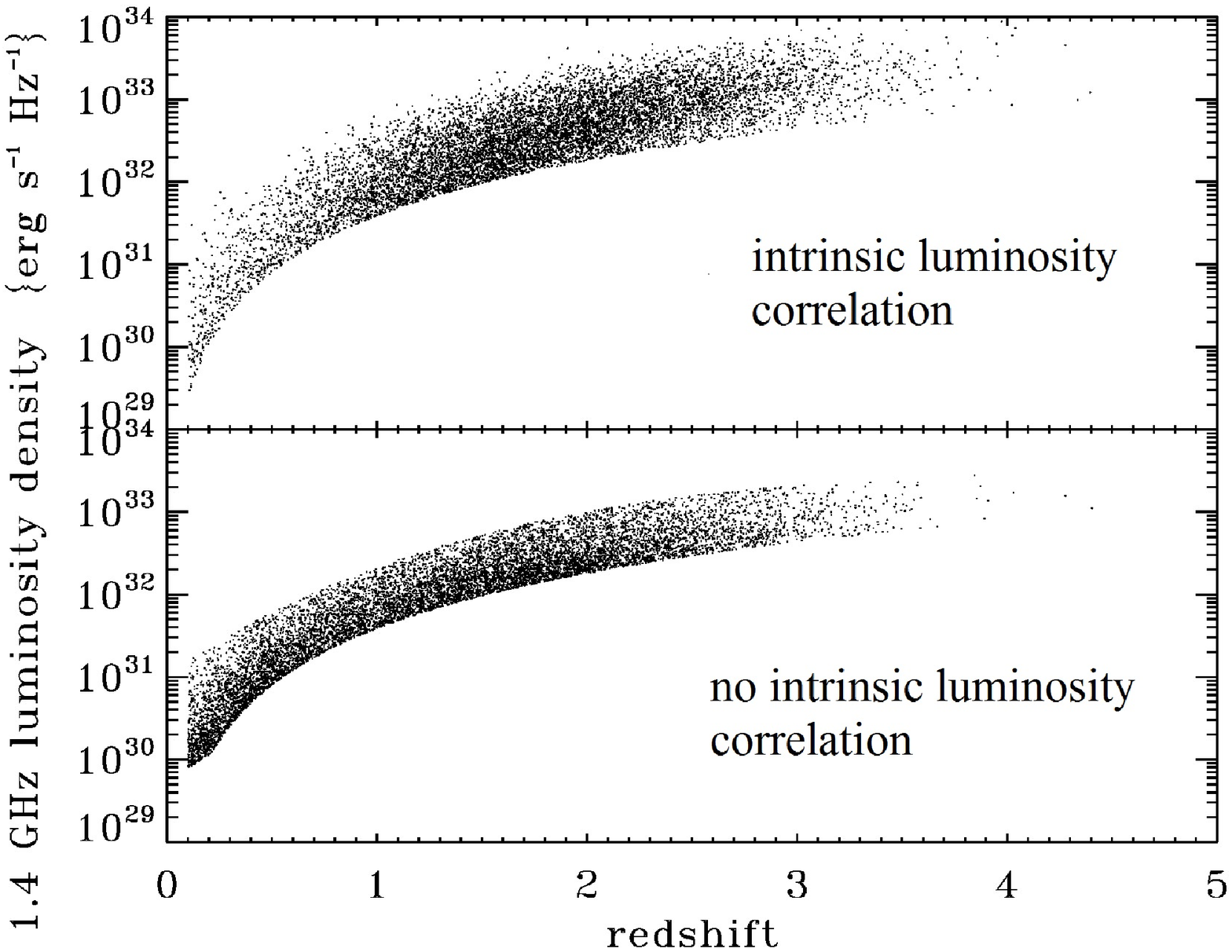}
\caption{The optical luminosities vs. redshift for the observed simulated data sets, for the case of intrinsically correlated ({\bf top} --- $\alpha=1.0$) and intrinsically uncorrelated ({\bf bottom} --- $\alpha=0$) radio and optical luminosities.  {\bf There is only one selection-induced truncation visible, that of the main curve at low luminosities which increases with redshift.  The apparent cutoff at high luminosities (and an additional, minor one for low luminosities which appears at very low redshifts for the uncorrelated case) is not an observational effect but rather an artifact of the underlying simulation and does not appreciably affect the analysis here, as discussed in \S \ref{intchar}.} }
\label{simopt}
\end{figure} 

\begin{figure}
\includegraphics[width=3.5in]{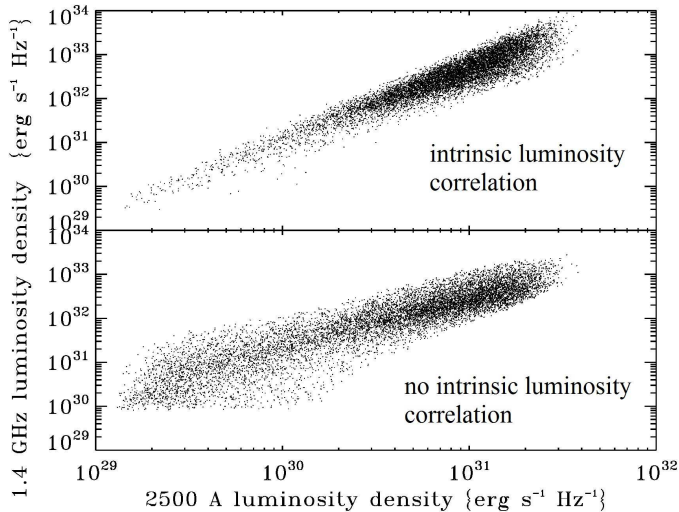}
\caption{The radio luminosities vs. redshift for the observed simulated data sets, for the case of intrinsically correlated ({\bf top} --- $\alpha=1.0$) and intrinsically uncorrelated ({\bf bottom} --- $\alpha=0$) radio and optical luminosities.  {\bf There is only one selection-induced truncation visible, that of the main curve at low luminosities which increases with redshift.  The apparent cutoff at high luminosities (and an additional, minor one for low luminosities which appears at very low redshifts for the uncorrelated case) is not an observational effect but rather an artifact of the underlying simulation and does not appreciably affect the analysis here, as discussed in \S \ref{intchar}. } }
\label{simrad}
\end{figure} 

In order to explore the effects of redshift evolutions and observational selection effects on population characteristics, in particular the $L$--$L$ correlation, we simulate populations with known intrinsic properties, such as LFs in two different wavebands, undergoing co-moving density and luminosity evolutions similar to those for observed AGNs. From this we then select an observed sample with two hypothetical flux-limits.  To develop and highlight comparisons with recently explored real populations \citep[e.g.][]{QP2,QP3} we have assumed this simulated population to be `quasars' observed by large area surveys and labeled the two wavebands `optical' and `radio,' but the conclusions as far as issues of $L$--$L$ correlations and population distributions are entirely general.  Our main goal is to see how well we recover the input characteristics, in particular the assumed $L$--$L$ correlation, using the methods we used in above papers and in \citet {CR1}.

\subsection{Simulated Population Characteristics} \label{intchar}

We have distributed the populations according to the following intrinsic characteristics,  now switching notation in some cases to  dimensionful luminosity $L_{a}$ for the luminosity in a given waveband and redshift $z$ rather than the dimensionless $x$ and $y$ and distance $r$ used in \S \ref{analytic}.  The populations have intrinsic ``local'' (that is before any redshift evolution effects are considered) differential LFs which obey a simple power law of the form
\begin{equation}
\psi_a\!(L'_{a}) = - {d\Phi(L'_{a}) \over dL'_{\rm a}}=\psi_{0,a} (L'_{a})^{\delta_a}\Theta(L'_a-L_{0,a}), 
\label{locallumfn}
\end{equation}
where $\Phi(L'_{a})$ is the cumulative local LF.  This results in a power-law local LF (with power-law index $\delta_a$) above an assumed minimum luminosity for the population $L_{0,a}$.  We then introduce luminosity evolution with redshift to the population with the functional form used for our AGN studies which has been shown to be a good fit to the data \citep{QP2,QP3}:
\begin{equation}
L_{a}(z)=L'_a \times g_a(z)\,\,\,\, {\rm with}\,\,\,\, g_a(z)={Z^{\rm k_a} \over 1+ (Z/Z_{\rm cr})^{\rm k_a}}
\label{eveq}
\end{equation}
where $Z \equiv 1+z$ as above, with potentially different parameters $(\delta_a, L_{0,a} , k_a)$ for each waveband.  This form allows for rapid evolution up to redshift $z_{\rm cr}$ then less rapid evolution at higher redshifts where rest-frame time changes are smaller.  The population also is simulated to have a co-moving density evolution $\rho(z)$ or the differential number evolution
\begin{equation}
{ {d\sigma(z)} \over {dz} }= \rho(z){dV\over dz}\propto e^{ {-(z-z_{\rm m})^2 } \over {2s} }
\label{deveq}
\end{equation}
where $\sigma(z)$ is the cumulative number evolution discussed above with $z_{\rm m}$ and $s$ as the mean redshift and variance. With the population characteristics distributed in this way, the overall LF in a waveband $a$ can be expressed as 
\begin{equation}
{\bar \Psi_{a}}\!(L_a,z) = \rho(z)\,\psi_a \left({L_{a} \over g_{a}\!(z)} \right)
/g_a\!(z),
\label{lumeq}
\end{equation}
This is because $\rho(z)$ quantifies the object density evolution with redshift, while what remains is the distribution over luminosities (which may change with redshift which is known as luminosity evolution).  To see that the latter is included, consider that at a particular redshift $z$ and particular luminosity $L_a$ one would be drawing from the local LF (a function of $L'_{a}=L_a  / g_a\!(z) )$ at a luminosity value that is lower (assuming positive luminosity evolution with redshift) by a factor of $1/g_a\!(z)$, and therefore drawing an appropriately higher corresponding LF value to account for the redshift evolution of the luminosity.  However to remain properly normalized so that integrating over all luminosities gives the total number of objects at any given redshift, the higher LF value must then be divided by the factor $g_a\!(z)$.  Mathematically, the latter can be seen by noting the relation, given the luminosity evolution, $dL'_a = dL_a / g_a\!(z)$.   

The total number of observed objects is then
\begin{equation}
N_{tot} = \int_0^{z_{max}} dz \int_{L_{\rm min}(z)}^\infty { dL_{a} \, \rho(z)
\, {dV \over dz} \, { {\psi_a\!\left(L_{a}/g_{a}\!(z)\right)} \over {g_a\!(z)} } } ,
\label{inteq}
\end{equation}
where the value of $L_{\rm min}(z)$ depends on the flux limit of the sample in waveband $a$.

In what follows we simulate a population in two different bands (which we will call optical and radio) with a simple power law intrinsic correlation between the local (prior to any redshift evolution) luminosities:
\begin{equation}
L'_{\rm rad} \propto (L'_{\rm opt})^{\alpha}
\label{corrdef}
\end{equation}
where $\alpha$ is the correlation index.  We explore the values $\alpha =0.0$ (i.e.~no correlation) and two different degrees of correlations with $\alpha=0.5$ and 1.0. 

For the luminosity evolutions, in order to span values approximately matching the intrinsic characteristics of real populations from previous analyses, we adopt the value $Z_{\rm cr}=3.7$ and $k_{\rm opt}=3.0$ and $k_{\rm rad}=4.5$.  For the LFs and density evolution, also to approximate intrinsic values seen in previous analyses, we adopt power law indexes, $\delta_{\rm opt}$=-2 and $\delta_{\rm rad}$=-2, and $z_m=2$ and $s$=0.75.   We also assume that the spectrum of sources in the short range of frequencies around each band can be approximated by a power law
\begin{equation}
L_a \propto \nu^{-\varepsilon_a}
\label{spectpow}
\end{equation}
with photon index values of $\varepsilon_{\rm opt}=0.5$ and $\varepsilon_{\rm rad}=0.4$.  We form Monte Carlo populations with these distributions by inverse transform sampling which allows random numbers to be generated uniformly on the interval [0,1] \citep[e.g.][]{Miller10}.   For concreteness we consider the optical luminosity density at 2500 \AA \, and the radio luminosity density at 1.4 GHz.  

{\bf The inverse transform sampling method requires assigning a lower and upper limit to the quantity being simulated, and the choice of these limits determines how many objects must be simulated in order to achieve a reasonable number of observed objects once the observational limits are imposed.  These limits result in an effective maximum luminosity (which varies with redshift because of the imposed luminosity evolution), and additionally an effective minimum luminosity at very low redshifts for the uncorrelated case, which can be seen visually in Figures \ref{simopt} and \ref{simrad}.  However because of the steepness of the LFs and the relative lack of very low redshift sources, these simulation cutoffs are much less significant for the observed data set than the main observational truncation at low luminosities --- i.e. there are far fewer objects in the $L$--$z$ regions of the former than the latter.  In the analyses discussed below in \S \ref{partcor} and \ref{ep} we treat only the observational truncation, so to the extent that we successfully recover the relevant intrinsic population characteristics it is done in spite of the presence of these additional and unaccounted for cutoffs, highlighting their sub-dominant nature and lack of significant effect on the analyses. }

\subsection{Simulated Selection Effects}

\begin{figure}
\includegraphics[width=3.5in]{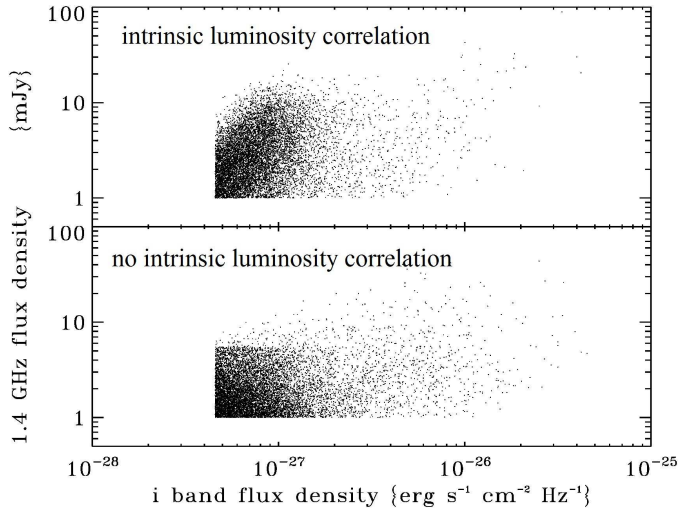}
\caption{The optical vs. radio luminosities for the observed simulated data sets, for the case of intrinsically correlated ({\bf top} --- $\alpha=1.0$) and intrinsically uncorrelated ({\bf bottom} --- $\alpha=0$) radio and optical luminosities.  As in Figure \ref{firstfig} it is clear that the {\it observed} luminosities can be correlated even if there is no intrinsic correlation between them. }
\label{simlumlum}
\end{figure} 

Because an optical observation is needed to identify a quasar via colors and provide a spectroscopic redshift, only those objects whose flux density is greater than the corresponding assumed minimum for detection in {\it both} wavebands is considered to be part of an observed sample.  With the populations simulated according to the intrinsic characteristics of \S \ref{intchar}, we then apply simulated flux-limited ``observations'' in both wavebands.  For simplicity, straightforwardness, and a connection to real data, the optical survey is taken to observe in a filter equivalent to the Sloan Digital Sky Survey (SDSS) $i$ band \citep[e.g.][]{SDSSQ} and have a universal magnitude limit of 19.1 in that band, and the radio survey is taken to be observing at 1.4 GHz have a universal flux density limit of 1 mJy.  The former is a simplified version of a limit that can be taken for the SDSS data release 7 quasar catalog \citep{SDSSQ} and the latter is a simplified version of the limit of the Faint Images of the Radio Sky at Twenty one centimeters (FIRST) survey \citep{FIRST1}.  Figure \ref{simfluxflux} shows the opt-radio flux-flux scatter diagram for two simulations; one for the case of no-correlation ($\alpha=0$) and one with
$\alpha=1$. These are the simulated ``observed" data we start with.

From the flux density of each object  object $j$ in waveband $a$, $f_{j,a}$, we calculate its luminosity density in that waveband with the luminosity-flux relation
\begin{equation}
f_{j,a} = { {L_{j,a} \, K_a(z)} \over {4 \, \pi \, D_L(z)^2  } }
\end{equation}
where $D_L(z)$ is the luminosity distance determined from the standard cosmology and $K_a(z)$ is the K-correction factor.  For a power law spectrum as in equation \ref{spectpow} the K-correction factor is 
\begin{equation}
K_a(z) = (1+z)^{1-\varepsilon_a}
\end{equation}
In Figures \ref{simopt}, \ref{simrad} we show the optical and radio luminosities vs. redshift, and in  \ref{simlumlum} we show radio luminosities vs. optical luminosities  for the 10,000 object ``observed'' simulated data sets.  As evident a strong observed correlation is present even for the uncorrelated ($\alpha=0$) sample.

\section{Analysis with Binned Partial Correlations} \label{partcor}

Here we explore the efficacy of determining correlations with data binned in redshift.  In the limit of infinitessimally narrow bins, the data within each bin should have no appreciable luminosity evolution and will be truncated parallel to the axes in the $L$--$L$ plane, and therefore the phenomena that induce $L$--$L$ correlations discussed in \S \ref{intro} will be irrelevant (select truncations parallel to the axes in the $L$--$L$ plane are shown in the top panel of Figure \ref{firstfig}).  Thus redshift binning has been used as a technique to deduce intrinsic $L$--$L$ correlations \citep[e.g.][]{Pavlidou12}.  The question still arises, however, whether analysis in finite-sized bins where these effects do not disappear completely is effective.  

A standard measure of partial correlations is the Pearson partial correlation coefficient \citep[PPCC --- e.g][]{RS07}, which expresses the partial correlation between two variables discounting their mutual dependence on a third:

\begin{equation}
r_{12,3} = {{r_{12} - r_{13}r_{23}} \over {[(1-r_{13}^{2})(1-r_{23}^{2})]^{1/2}}}
\label{pearsoneq}
\end{equation}
where $r_{ab}$ is the standard sample Pearson's moment correlation (PMC  --- commonly known as Pearson's $r$) between variables $a$ and $b$
\begin{equation} \label{pmc}
r_{ab} = \frac{\sum_{i} (a_{i}- \overline{a}) (b_{i}- \overline{b})}{N \sigma_a \sigma_b}
\end{equation}
where $\sigma_a = \sum \sqrt{\frac{1}{N}(a_{i}- \overline{a})^2}$ is the standard deviation of the $a$ values and $N$ is the total number of data points. 

It is important to note that the PMC and PPCC are measures of the extent to which two variables are correlated, in the sense of being related by some function.  However, they do not shed any light on the nature of the correlation function itself, and a higher value does not necessarily indicate a steeper correlation function (or vice-versa), only that the data more closely adhere to the function whatever it may be.   In this work we calculate PMCs and PPCCs using the {\it logarithm} of the luminosity values (and linear redshift values), in order to reduce the potential outsize effect of a small number of objects with a very high luminosity in a given bin.

We bin the data and then examine a) the two luminosity-redshift correlations, b) the $L$--$L$ correlation, and c) the {\it partial} $L$--$L$ correlation for two cases: i) the raw observed luminosities, and ii) the so-called ``local'' luminosities with the best-fit redshift evolution removed.  The differences between the $L$--$L$ full and partial correlations between the two cases can reveal how much of the $L$--$L$ correlation is physically real and how much is due to redshift evolution.  

The most effective binning method for our needs, taking into consideration the data that we deal with, was found to be an equal number of objects per bin since objects are distributed unevenly across redshift. If we divide bins instead with uniform redshift size per bin, the few highest redshift bins end up with too few objects, resulting in unrealistic, erratic, and unreliable correlation coefficients for these bins. The number of objects in the least populated bins could be increased by increasing the width of the bins in redshift, but this leads to severely flux-limit induced correlations as discussed above.  On the flip side, having an equal number of objects per bin and many bins would lead to bins with excessively small redshift ranges due to a high number of objects at those redshifts. While this does not make the $L$--$L$ correlations unreliable, it does hide intrinsic luminosity-redshift correlations since the redshift range is too small to detect redshift dependent correlations. The optimum number of bins is thus the result of a trade-off between having some of the bins be too narrow and some too wide, and depends on the size of the data set.

The effects of cosmological evolutions and observational selection biases can be can be investigated by examining raw as well as local luminosities for the uncorrelated and correlated simulated data sets discussed in \S \ref{sims}.  

\begin{figure}
\includegraphics[width=3.5in]{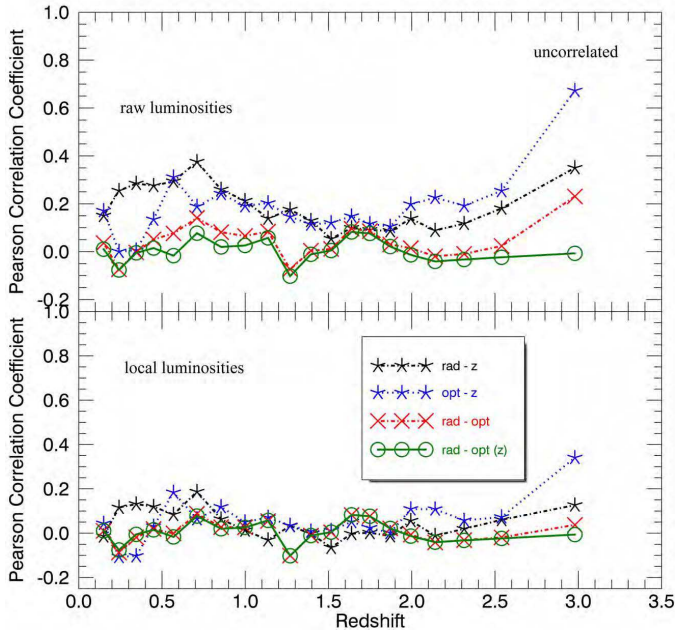}
\caption{Radio-redshift, optical-redshift, and radio-optical PMCs, and radio-optical partial with redshift PPCCs in 20 bins of redshift with an equal number of objects per bin for raw (\textbf{top}) and local (\textbf{bottom}) luminosities for the intrinsically uncorrelated simulated observed radio and optical quasar data.  Points are plotted at the average redshift and correlation values for each bin.  We see that the lack of correlation between the two luminosities is manifest, and that the luminosity-redshift correlations present in the raw luminosities are removed when considering the local luminosities. }
\label{ucsim20}
\end{figure}

Figure \ref{ucsim20} shows intrinsically uncorrelated simulated radio-optical data in 20 bins of redshift for both raw (top panel) and local luminosities (bottom panel). As expected, the radio-optical partial correlation coefficients for both raw and local luminosities are all approximately zero since this simulated data was designed to have no intrinsic correlation between the optical and radio luminosities. As hypothesized, in the top panel of Figure \ref{ucsim20}, we can see the radio-redshift and optical-redshift correlation coefficients to be non-zero since the population has luminosity evolution. Since we are not using infinitesimally small redshift bins, there is an automatic influence of the flux-limit on the luminosities vis-a-vis redshift, which further contributes to a higher radio-redshift and optical-redshift correlation. Moreover, the radio-optical full correlation coefficients can be observed to be relatively higher than the partial correlation coefficients because the former are not disregarding their mutual dependence on redshift. This plot also demonstrates the contrast between using large versus small bins. The last bin (at around average redshift of three, with the largest  redshift range) of both panels of Figure \ref{ucsim20} shows a relatively strong dependence
of the luminosities on redshift,  and as a result  has a higher radio-optical raw luminosity full correlation as well. This behavior is expected and is due to two reasons; one, as we discussed earlier, having larger redshift ranges brings in the flux-limit effect into the luminosity dependence on redshift, automatically and misleadingly strengthening the correlation between luminosities and redshift; and two, having a larger redshift range shows a relatively large artificial PMC correlation, that disappears using the partial correlation PPCC.

\begin{figure}  
\includegraphics[width=3.5in]{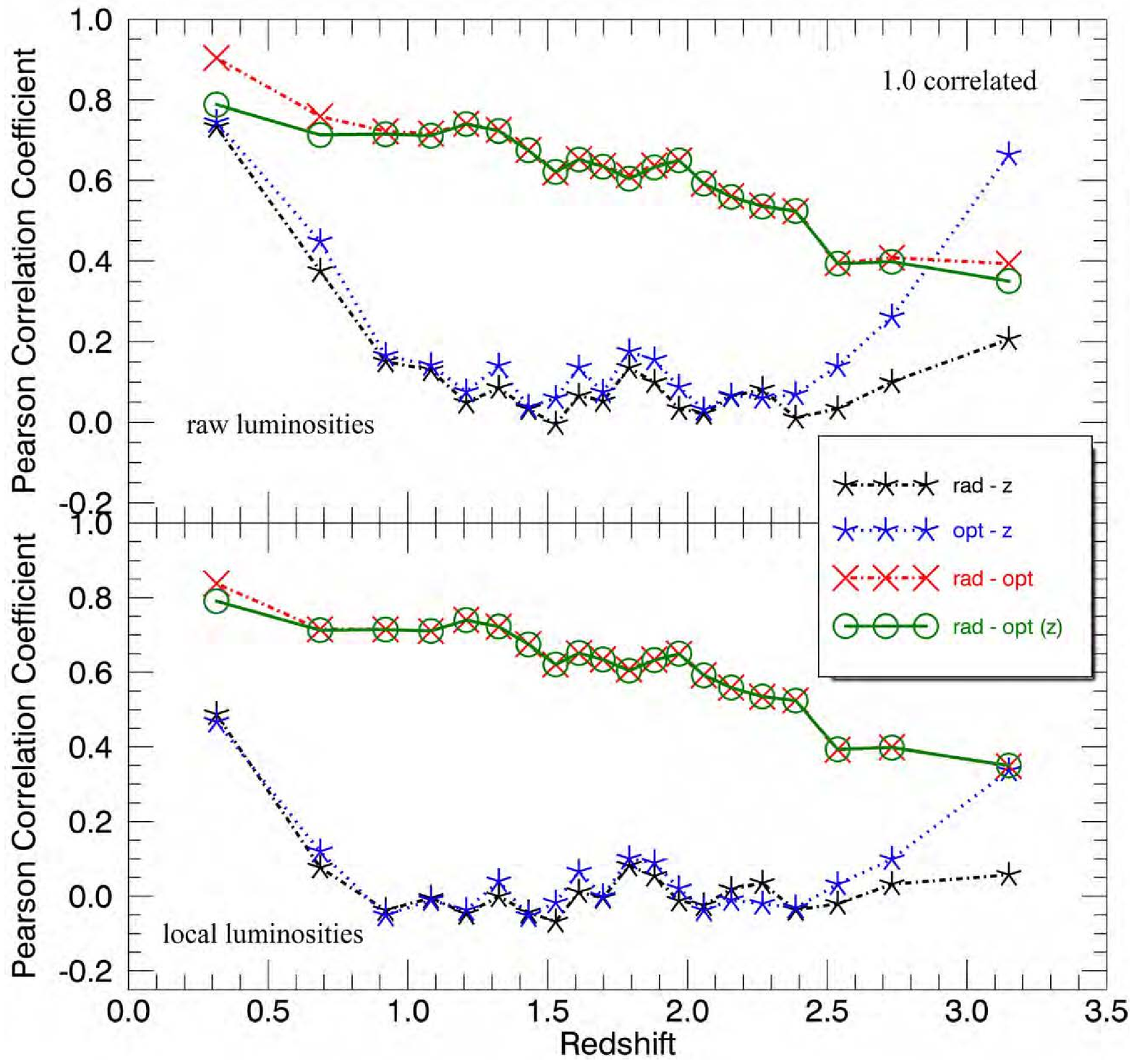}
\caption{Radio-redshift, optical-redshift, and radio-optical PMCs, and radio-optical partial with redshift PPCCs in 20 bins of redshift with an equal number of objects per bin for raw (\textbf{top}) and local (\textbf{bottom}) luminosities for the intrinsically 1.0-correlated simulated observed radio and optical quasar data.  Points are plotted at the average redshift and correlation values for each bin.  We see that the strong correlation between the two luminosities is manifest, that the luminosity-redshift correlations present in the raw luminosities are removed when considering the local luminosities, and that removing the luminosity-redshift correlations decreases the divergence present in some bins between the PMCs and PPCCs for the luminosity-luminosity correlations.}
\label{csim20}
\end{figure}

\begin{figure} 
\includegraphics[width=3.5in]{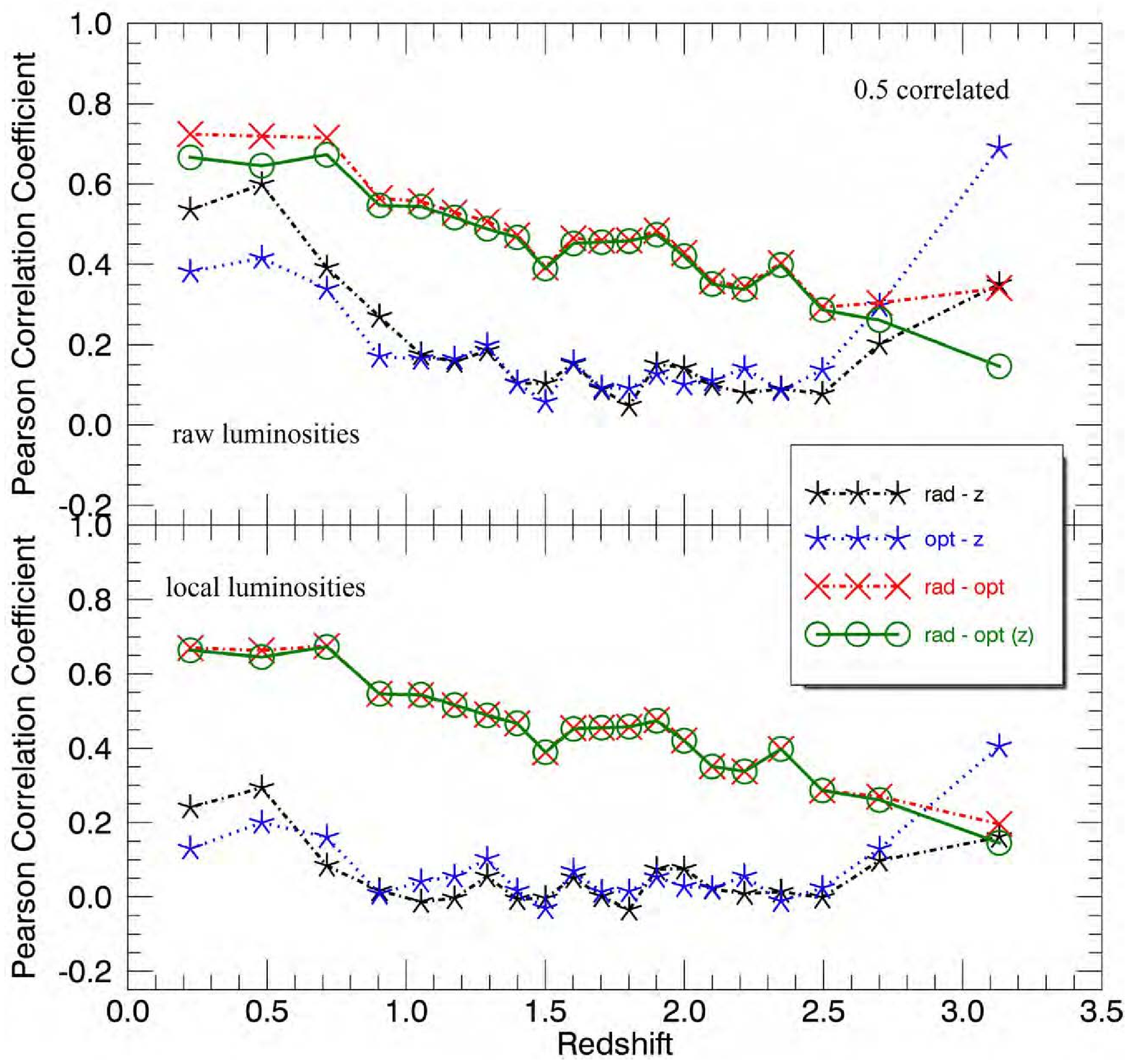}
\caption{Radio-redshift, optical-redshift, and radio-optical PMCs, and radio-optical partial with redshift PPCCs in 20 bins of redshift with an equal number of objects per bin for raw (\textbf{top}) and local (\textbf{bottom}) luminosities for the intrinsically 0.5-correlated simulated observed radio and optical quasar data.  Points are plotted at the average redshift and correlation values for each bin.  We see that the strong correlation between the two luminosities is manifest, that the luminosity-redshift correlations present in the raw luminosities are removed when considering the local luminosities, and that removing the luminosity-redshift correlations decreases the divergence present in some bins between the PMCs and PPCCs for the luminosity-luminosity correlations.}
\label{scsim20}
\end{figure}

In comparison, we expected luminosity dependence on redshift and the radio-optical correlations to be smaller for the local luminosities as shown in the bottom panel of Figure \ref{ucsim20}. The utilization of local luminosities removes the redshift evolution, allowing us to observe correlations that exist sans redshift dependence.  This is exactly what is seen, as the full radio-optical and the partial radio-optical correlation coefficients align almost perfectly with each other in the bottom panel of Figure \ref{ucsim20}. However, local luminosities still do not remove the effect of the flux-limit, which is why we do not see a completely non-existent redshift dependence in luminosities, and which is why the last bin still has a relatively higher luminosity dependence on redshift than the other bins.

Figures \ref{csim20} and \ref{scsim20} show the cases of intrinsically correlated simulated radio-optical data, with the correlation power law index (c.f. equation \ref{corrdef}) $\alpha$=1.0 and 0.5, respectively, in 20 bins of redshift for both raw (top panel) and local luminosities (bottom panel).  Compared to the uncorrelated case, these manifest some distinctly contrasting features. As anticipated, the radio-optical partial correlation coefficients for both the top and bottom panels of Figure \ref{csim20} are higher than for Figure \ref{scsim20} and both are much higher than in Figure \ref{ucsim20} since the simulated data was designed to have intrinsic correlation between the luminosities. As in Figure \ref{ucsim20}, the luminosity-redshift correlations are generally non-zero in the top panels drop lower (almost to zero) in the bottom panel of Figures \ref{csim20} and \ref{scsim20} since using local luminosities removes their intrinsic dependence on redshift.  

We see from considering the above figures that full and partial correlation analysis in appropriately sized bins of redshift is a useful tool for determining presence or lack of, and at least qualitiatively the degree, of intrinisic correlation between luminosities in a doubly flux-limited sample.  The simulated data sets with intrinsic correlations between the luminosities (whether the actual form of the correlation is linear or sub-linear) show significantly higher partial correlations between the luminosities and in particular the local luminosities than is the case for the simulated data set with no intrinsic correlation between the luminosities, which manifests nearly zero average partial correlation between the local luminosities.

\section{Demonstration of Non-Paramatric Techniques With Simulated Data Sets} \label{ep}

In recent works \citep{QP1,BP1,BP2,QP2,BP3,QP3} we used multiwavelength extensions of methods first proposed by Efron and Petrosian \citep{EP92,EP99} to recover the intrinsic distributions and correlations of the luminosities and redshifts in flux-limited multiwavelength data.  Here we apply these techniques to the simulated data sets developed in \S \ref{sims} to demonstrate how well we recover the input characteristics.

\subsection{Luminosity Evolutions} \label{rev}

\begin{figure} 
\includegraphics[width=3.5in]{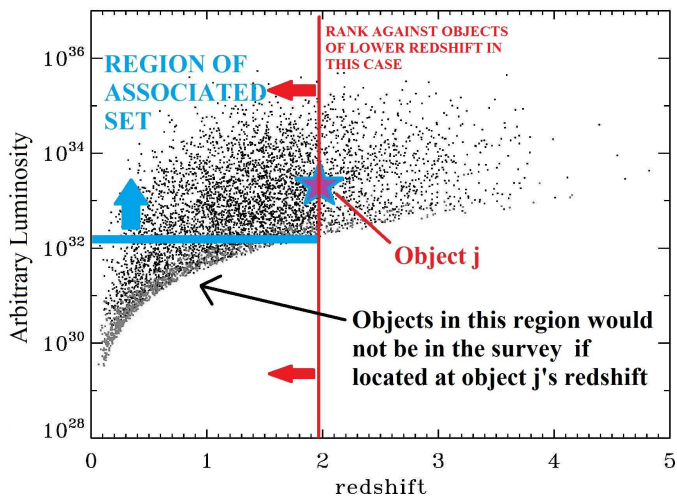}
\caption{Depiction of the associated set for a particular object in a hypethetical single-flux-limited single waveband survey.  Associated sets are introduced in \S \ref{rev}. }
\label{assfig}
\end{figure}

\begin{figure}
\includegraphics[width=3.5in]{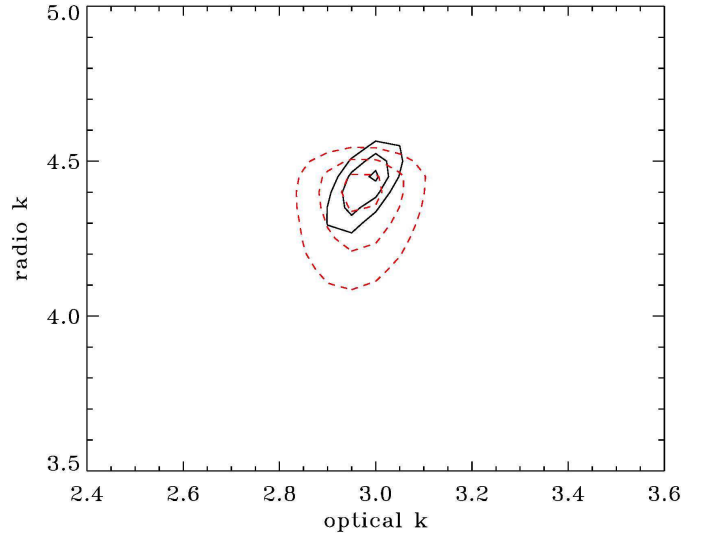}
\caption{The $1\sigma$, $2 \sigma$, and $3 \sigma$ contours for the simultaneous best fit values of $k_{\rm opt}$ and $k_{\rm rad}$ of the simulated samples, for the forms of the luminosity evolutions given by equation \ref{eveq}, and for simulations with intrinsic (solid) and no intrinsic (dashed red) correlations between the luminosities.  It is seen that the input intrinsic luminosity evolutions ($k_{\rm opt}=3.0$ and $k_{\rm rad}=4.5$ --- see equation \ref{eveq} in \S \ref{intchar}) are recovered to within small deviations.  }
\label{alphas}
\end{figure}

We determine the correlations between luminosity and redshift by using a variant of a rank test statistic modified with the use of {\it associated sets} which are unbiased sets for comparison.  The test statistic
\begin{equation}
\tau = {{\sum_{j}{(\mathcal{R}_j-\mathcal{E}_j)}} \over {\sqrt{\sum_j{\mathcal{V}_j}}}}
\label{tauen}
\end{equation}
tests the independence of two variables in a dataset, say ($x_j,y_j$) for  $j=1, \dots, n$.  Here $\mathcal{R}_j$ is the dependent variable ($y$) rank of the data point $j$ in a set associated with it, $\mathcal{E}_j=(1/2)(n+1)$ is the expectation value and $\mathcal{V}_j=(1/12)(n^{2}+1)$ is the variance, where $n$ is the number of objects in object $j$'s associated set.  For untruncated data (i.e. data truncated parallel to the axes) the set associated with point $j$ includes all of the points with a lower (or higher, but not both) independent variable value ($x_k < x_j$).  If the data is truncated one must form the {\it associated set} consisting only of those points of lower (or higher, but not both) independent variable ($x$) value {\it that would have been observed if they were at the $x$ value of point $j$ given the truncation}.  Figure \ref{assfig} shows a graphical description of an associated set for a given example data point. 

If ($x_j,y_j$) are independent then the ranks $\mathcal{R}_j$ should be distributed randomly and $\tau$ should sum to near zero.  Independence is rejected at the $m \, \sigma$ level if $\vert \, \tau \, \vert > m$.  To find the best fit correlation bewteen $y$ and $x$ the $y$ data are adjusted by defining $y'_j=y_j/F(x_j)$  and the rank test is repeated, with different values of parameters of the function $F$ until $y'$ and $x$ are determined to be uncorrelated.

In the case here of multiband luminosity and redshift data, for determining the redshift evolution of luminosity we can treat redshift as the independent variable and the luminosities as dependent variables.  The problem becomes one of determining the evolution factors $k_a(z)$ in the functions $g_a(z)$ in equation \ref{eveq} which render each luminosity uncorrelated with redshift.  In the three dimensional case, properly taking into account the data truncations is important because we now are dealing with a three dimensional distribution  ($L_{\rm rad}, L_{\rm opt}, z$) and two correlation functions ($g_{\rm rad}\!(z)$ and $g_{\rm opt}\!(z)$), plus we can find the true intrinsic correlation in this case because the truncation effects in the luminosity-redshift space are known and redshift is the independent variable in both cases.

Since we have two criteria for truncation, the associated set for each object $k$ includes only those objects that are sufficiently luminous in both bands to have been in the survey if they were located at the redshift of the object in question.  The luminosity cutoff limits for a given redshift must also be adjusted by factors of $g_{\rm opt}\!(z)$ and $g_{\rm rad}\!(z)$. Consequently, we have a two dimensional minimization problem, because objects will drop in and out of associated sets as $g_{\rm opt}\!(z)$ and $g_{\rm rad}\!(z)$ change, leading to changes in the calculated ranks in equation \ref{tauen}.

We form a test statistic $\tau_{\rm comb} = \sqrt{\tau_{\rm opt}^2 + \tau_{\rm rad}^2} $ where $\tau_{\rm opt}$ and $\tau_{\rm rad}$ are those evaluated considering the objects' optical and mid-infrared luminosities, respectively.  The favored values of $k_{\rm opt}$ and $k_{\rm rad}$ are those that simultaneously give the lowest $\tau_{\rm comb}$ and, again, we take the $1 \sigma$ limits as those in which $\tau_{\rm comb} \, < 1$.  Figure \ref{alphas} shows the 1 and 2 $\sigma$ contours for $\tau_{\rm comb}$ as a function of $k_{\rm opt}$ and $k_{\rm rad}$ for the simulated data sets.  We see that the input intrinsic luminosity evolutions are recovered.  We note here that we were able to recover the input intrinsic luminosity evolutions in the case of the intrinsically correlated luminosities without consideration of an orthogonal ``correlation reduced'' radio luminosity as explored in previous works \citep[e.g][]{QP2}.  

\subsection{Density Evolution}\label{dev}

\begin{figure}
\includegraphics[width=3.5in]{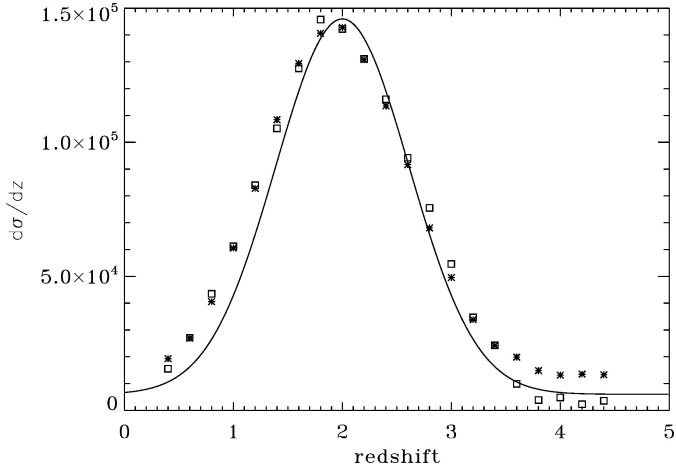}
\caption{The differential density function $d \sigma\!(z)/dz$ vs. redshift determined from the simulated data sets calculated as in \S \ref{dev}, for the cases of intrinsic (stars) and no intrinsic (squares) correlations between the luminosities.  The normalization of $d \sigma\!(z)/dz$ here is arbitrary.  It is seen that the input intrinsic redshift distribution of the population ($z_m=2.0$, $s=0.75$ --- see equation \ref{deveq} in \S \ref{intchar}) is relatively closely recovered.  For reference a Gaussian with these input characteristics is also plotted.  }
\label{densev}
\end{figure}

The cumulative density function $\sigma(z)$ and differential density function $\rho\!(z)$ are related as in equation \ref{ssigeq}.  Expressed as a function of $z$ instead of distance this would give for the cumulative number of objects at redshifts less than $z$
\begin{equation}
\sigma(z) = \int_0^z { {{dV} \over {dz}} \, \rho\!(z) \, dz}
\end{equation}  
Following \citet{P92} based on the method of \citet{L-B71} which is equivalent to a maximum likelihood estimate, $\sigma(z)$ can be calculated by

\begin{equation}
\sigma(z) = \prod_{j}{(1 + {1 \over m(j)})}
\label{sigmaeqn}
\end{equation}
where the set of $j$ includes all objects with a redshift lower than or equal to $z$, and $m(j)$ is the number of objects with a redshift lower than the redshift of the object at redshift $z$ {\it which are in that object's associated set}.  In this case, the associated set is again those objects with sufficient optical and radio luminosity that they would be seen if they were at redshift $z$.  The use of only the associated set for each object accounts for the biases introduced by the data truncation. 

However, to determine the density evolution, the luminosity evolution determined in \S \ref{rev} must be taken out.  Thus, the objects' optical and radio luminosities, as well as the optical and radio luminosity limits for inclusion in the associated set for given redshifts are scaled by taking out factors of $g_{\rm opt}\!(z)$ and $g_{\rm rad}\!(z)$ which are determined as above.  The preceding method is fully adequate if there is a uniform selection function across redshift for quasars at a given flux.  The differential density evolution $d \sigma\!(z)/dz$ is shown in Figure \ref{densev}.  It is seen that the input intrinsic redshift distribution of the population is recovered.

\subsection{Local luminosity functions}

We first obtain a cumulative local luminosity function
\begin{equation}
\Phi_a\!(L_{a}') = \int_{L_{a}'}^{\infty} {\psi_a\!(L_{a}'') \, dL_{a}''}
\end{equation}
which, following \citet{P92} using the method of \citet{L-B71}, $\Phi_a\!(L_{a}')$, can be calculated by
\begin{equation}
\Phi_a\!(L_{a}') = \prod_{k}{(1 + {1 \over n(k)})}
\label{phieq}
\end{equation}
where $k$ runs over all objects with a luminosity greater than or equal to $L_a$, and $n(k)$ is the number of objects with a luminosity higher than the luminosity of object $k$ which are in object $k$'s associated set, which in this case consists of those objects which would be in the survey if they were at object $k$'s luminosity considering the luminosity limits for inclusion at each object's given redshift in both optical and radio. 
The local LF $\psi_a\!(L_{a}')$ is 
\begin{equation}
\psi_a\!(L_{a}') = - {d \Phi_a\!(L_{a}') \over dL_{a}'}
\label{psieqn}
\end{equation}

In \S \ref{rev} we determined the luminosity evolutions for the optical and radio luminosities.  We can form the local optical $\psi_{\rm opt}\!(L_{\rm opt}')$ and radio $\psi_{\rm rad}\!(L_{\rm rad}')$ LFs straightforwardly, by taking the evolutions out.  As before, the objects' luminosities, as well as the luminosity limits for inclusion in the associated set for given redshifts, are scaled by taking out factors of $g_{\rm rad}\!(z)$ and $g_{\rm opt}\!(z)$, with $k_{\rm rad}$ and $k_{\rm opt}$ determined in \S \ref{rev}.  Alternate methods of determining the local LF exist, such as binning objects in redshift and constructing the LF with the $1/V_{\rm max}$ method, as in \citet{PC00}.   However using the method here with local de-evolved luminosities has the advantage of using all of the objects to construct the local LF.

\begin{figure} 
\includegraphics[width=3.5in]{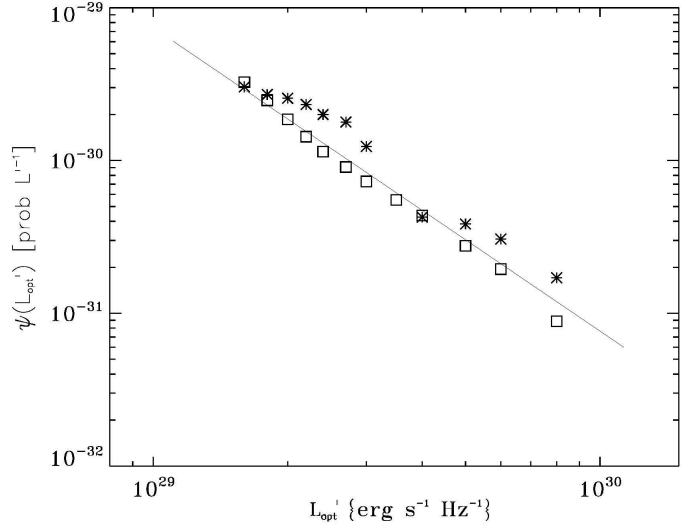}
\caption{The local optical LF $\psi_{\rm opt}\!(L_{\rm opt}')$ for the simulated data sets, for the for the cases of intrinsic (stars) and no intrinsic (squares) correlations between the luminosities.  It is seen that the input intrinsic local luminosity distributions of the populations ($\delta_{\rm opt}=2.0$ --- see equation \ref{locallumfn} in \S \ref{intchar}) are recovered.  For reference a line indicating a power law slope of $\delta_{\rm opt}=2.0$ is shown.}
\label{psiopt}
\end{figure}

\begin{figure} 
\includegraphics[width=3.5in]{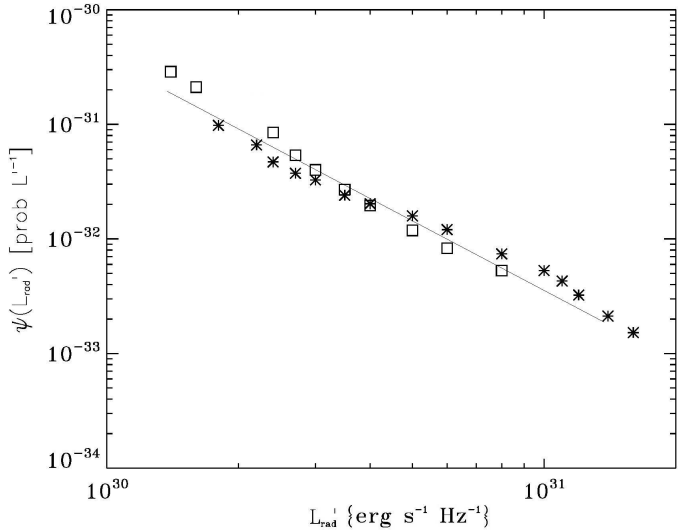}
\caption{The local radio LF $\psi_{\rm rad}\!(L_{\rm rad}')$ for the simulated data sets, for the for the cases of intrinsic (stars) and no intrinsic (squares) correlations between the luminosities.  It is seen that the input intrinsic local luminosity distributions of the populations ($\delta_{\rm rad}=2.0$ --- see equation \ref{locallumfn} in \S \ref{intchar}) are recovered.  For reference a line indicating a power law slope of $\delta_{\rm rad}=2.0$ is shown.  }
\label{psirad}
\end{figure}

Figures \ref{psiopt} and \ref{psirad} show the local differential $\psi_{\rm opt}\!(L_{\rm opt}')$ optical and radio LFs respectively determined for the simulated data sets.  Here we obtain the derivative of $\Phi_a\!(L_{a}')$ by fitting a simple cubic spline interpolation to $\Phi_a\!(L_{a}')$ and taking the derivative at various points where the spline is well behaved.  We see that we recover the input intrinsic LFs.

\subsection{Intrinsic Luminosity-Luminosity Correlations} \label{bff}

Having determined that binned PPCCs provide a potentially reliable method of determining the presence of an intrinsic correlation between luminosities, we use the technique introduced in \citet{CR1} to extract the best-fit power-law form of that correlation assuming that the luminosity evolutions have been determined as in \S \ref{rev}.  We perform a variable transformation by defining the so-called ``correlation reduced'' local luminosity as:

\begin{equation}
L'_{\rm crr} = {  {L'_{a} } \over { \left( {{ L'_{\rm opt} } \over {L_{\rm fid}} } \right)^{\alpha} } }
\label{deltaeq}
\end{equation}  
where $L_{\rm fid}$ is some fiducial luminosity to avoid exponentiating a dimensioned number, whose actual value is irrelevant.  Then for a range of values of $\delta$ we compute the median value of the PPCC between $L'_{\rm crr}$ and $L'_{\rm opt}$ in bins.  The value of $\alpha$ that results in a median PPCC of zero is the best-fit value for the power law exponent for the intrinsic correlation between $L'_{a}$ and $L'_{\rm opt}$.  

\begin{figure}
\hspace*{-0.1in} 
\includegraphics[width=3.5in]{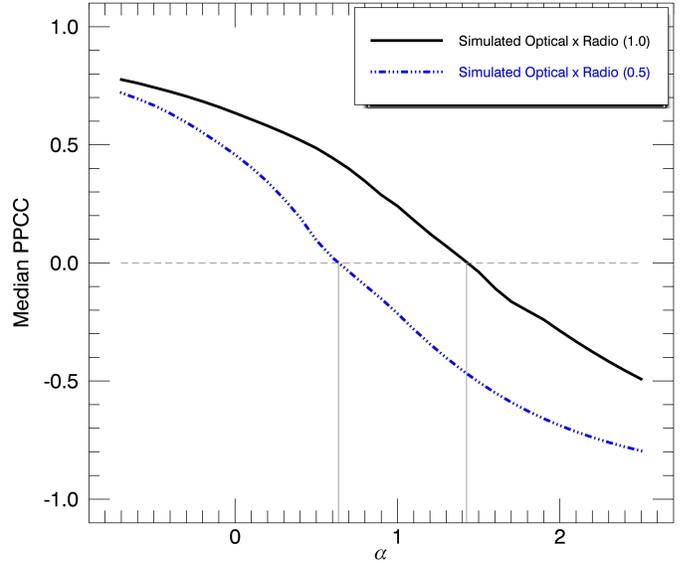}
\caption{Median of binned PPCC values for the ``correlation reduced local luminosity'' (see equation \ref {deltaeq}) versus local optical luminosity for the cases of the (1) intrinsically 1.0 and (2) 0.5-correlated simulated observed radio and optical quasar data, using 20 equally populated redshift bins.  The best-fit correlation between the luminosities is the $\alpha$ value that gives a median PPCC value of zero, as discussed in \S \ref{bff}.  The best-fit $L$--$L$ correlation power-law index values are $\alpha=$1.41$\pm$0.1 for the 1.0 correlated case and $\alpha=$0.65$\pm$0.1 for the 0.5 correlated case.  We see that this technique is somewhat reliable for recovering the known power-law form of the intrinsic correlation in the simulated data.  }
\label{deltas}
\end{figure}

Figure \ref{deltas} shows such a median PPCC vs. $\alpha$ for the intrinsically 1.0 (black)- 0.5(blue) -correlated simulated observed radio and optical quasar data, along with those for two real data sets discussed in \S \ref{real}.   The 1$\sigma$ range of uncertainties reported for these values is determined by considering the $\chi^2$ vs. $\alpha$ distribution.  We see that this technique recovers quite well the known power-law form of the intrinsic correlation in the 0.5-correlated simulated data, and somewhat overestimates the power-law value in the case of the 1.0-correlated simulated data.

\section{Demonstration: Luminosity-Luminosity Correlations in Real Quasars}\label{real}

We now show a partial correlation analysis and determination of the $L$--$L$ correlation index $\alpha$ of equation \ref{deltaeq} with two real observed two-flux-limited quasar data sets: (1) a set of optical and radio luminosities used in \citet{QP2}, and (2) a set of optical and mid-infrared luminosities used in \citet{QP3}.  Part of this analysis applied to the radio-optical data set was previously demonstrated in the conference proceeding \citet{CR1}.

The best-fit redshift evolutions for the luminosities of the form of equation \ref{eveq} are determined \citet{QP2} and \citet{QP3} respectively, with methods verified here in \S \ref{rev}.

\begin{figure} 
\includegraphics[width=3.5in]{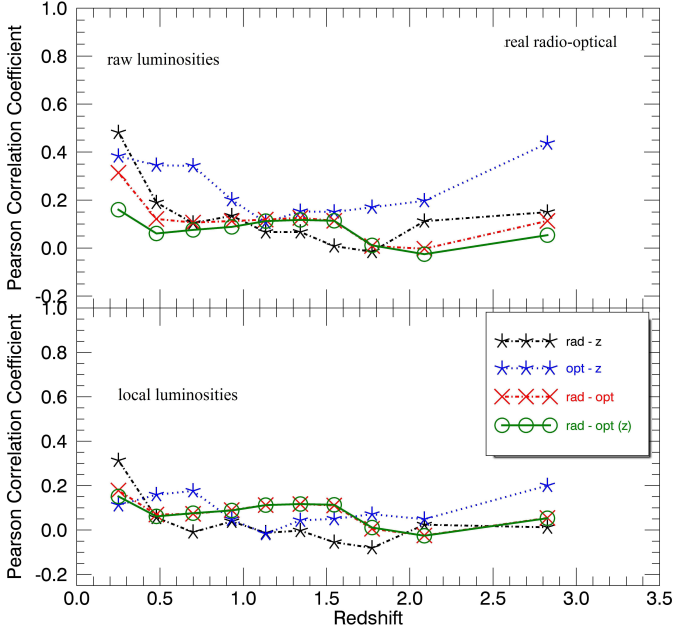}
\caption{Radio-redshift, optical-redshift, and radio-optical PMCs, and radio-optical partial with redshift PPCCs in ten bins of redshift with an equal number of objects per bin for raw (\textbf{top}) and local (\textbf{bottom}) luminosities for the real observed radio and optical quasar data from \citet{QP2}.  Points are plotted at the average redshift and correlation values for each bin.  We see that the two luminosities are only moderately intrinsically correlated, as the PPCCs are all positive but small. }
\label{optrad10}
\end{figure}

\begin{figure}
\hspace*{-0.1in} 
\includegraphics[width=3.5in]{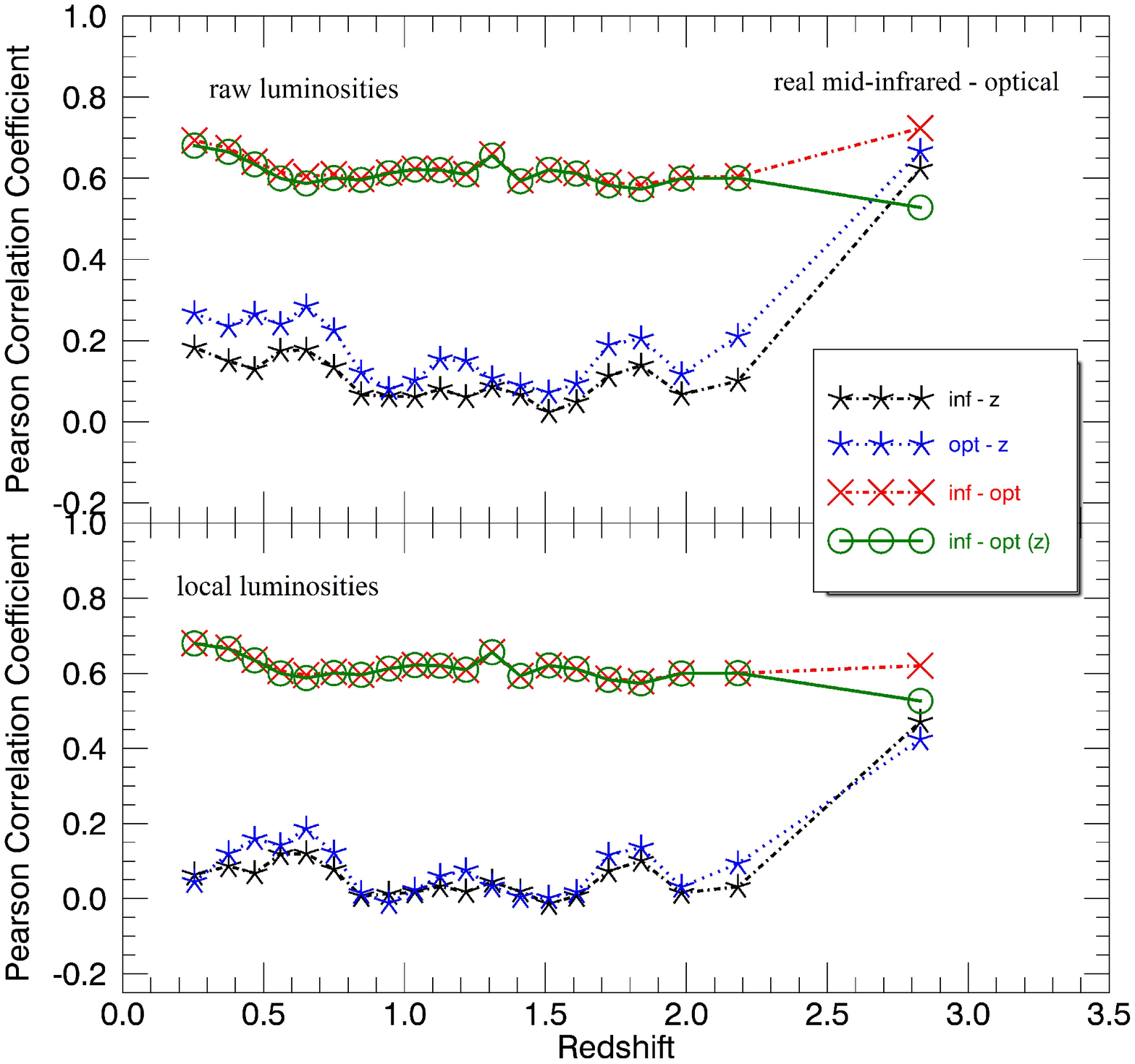}
\caption{Infrared-redshift, optical-redshift, and infrared-optical PMCs, and infrared-optical partial with redshift PPCCs in 20 bins of redshift with an equal number of objects per bin for raw (\textbf{top}) and local (\textbf{bottom}) luminosities for real observed mid-infrared and optical quasar data from \citet{QP3}.  Points are plotted at the average redshift and correlation values for each bin.  We see that the two luminosities are very strongly intrinsically correlated.}
\label{optinf20}
\end{figure}

\begin{figure}
\hspace*{-0.1in} 
\includegraphics[width=3.5in]{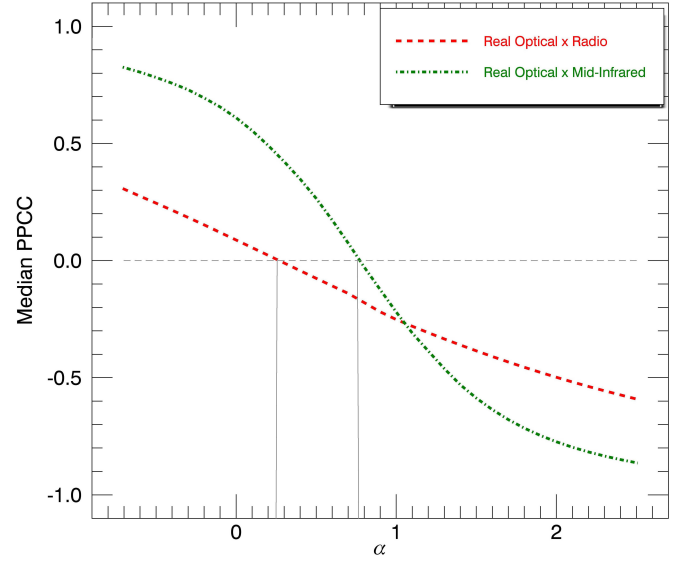}
\caption{Median of binned PPCC values for the ``correlation reduced local luminosity'' (see equation \ref {deltaeq}) versus local optical luminosity for the cases of the (1) the real radio and optical and (2) real mid-infrared and optical quasar data.  10 and 20 equally populated redshift bins were used for the former and the latter respectively.  The best-fit correlation between the luminosities is the $\alpha$ value that gives a median PPCC value of zero, as discussed in \S \ref{bff}.  The best-fit $L$--$L$ correlation power-law index values are $\alpha=$0.25$\pm$0.15 for raido-optical and $\alpha=$0.75$\pm$0.1 for mid-infrared-optical.  }
\label{deltas2}
\end{figure}

Figure \ref{optrad10} shows the PMCs and PPCCs for the optical-radio data set with ten bins of redshift for both raw (top panel) and local luminosities (bottom panel).  As this data set is quite a bit smaller than any of the simulated data sets considered above or the optical-mid-infrared data set, a smaller number of bins is warranted as discussed in \S \ref{partcor}.  As can be seen there, the radio-optical PPCCs are small yet not insignificant, with only two bins exhibiting radio-optical PPCCs equal to or less than zero. The radio-optical PMCs align almost perfectly with the PPCCs in the bottom panel of Figure \ref{optrad10}, indicating that removing the redshift evolution removes almost all of the excess induced correlation between the luminosities. The radio-optical PPCCs maintain their magnitudes across both the panels of Figure \ref{optrad10}, providing us a fairly reliable confirmation of the small yet not insignificant correlation between radio and optical luminosities.

Figure \ref{optinf20} shows real mid-infrared and optical data in 20 bins of redshift for both raw (top panel) and local luminosities (bottom panel). Figure \ref{optinf20} can be observed to clearly have features quite similar to the case of intrinsically correlated simulated data --- in particular the high PMCs and PPCCs which don't diminish significantly for the local luminosities, as shown in Figure \ref{csim20}. Figure \ref{optinf20} displays high $L$--$L$ PMC and PPCC values across all bins, signifying a high intrinsic correlation between mid-infrared and optical radiation being emitted by the observed quasars. Once we remove the intrinsic redshift-evolution of the luminosities and use local luminosities (bottom panel of Figure \ref{optinf20}), the infrared-optical PMCs drop slightly compared to the case of raw luminosities (top panel of Figure \ref{optinf20}) and align almost perfectly with the PPCCs. This indicates that the non-intrinsic, flux-limit induced redshift dependence of the luminosities is almost negligible in all but the highest redshift bin, where even in the bottom panel the full infrared-optical PMC of the highest redshift bin is larger than the PPCC. This anomalous behavior predictably signifies that the highest redshift bin still has a non-intrinsic redshift dependence of luminosities owing to the relatively larger redshift range.  These results indicate that the mid-infrared and optical luminosities are highly intrinsically correlated whereas the radio and optical luminosities characterize a much smaller, although still present, intrinsic correlation.  

We can then apply the techniques of \S \ref{bff} to determine the best-fit power-law form of the intrinsic correlations between the radio and optical and mid-infrared and optical luminosities.  These are shown in Figure \ref{deltas2}.  The results favor a higher power law of intrinsic correlation between the mid-infrared and optical luminosities (favoring a $\alpha \sim 0.7$) than for the radio and optical luminosities (favoring a $\alpha \sim 0.2$). We briefly discuss the physical implications of this in \S \ref{disc}.

\section{Summary and Discussion} \label{disc}

Understanding the true correlation between luminosities in different wavebands is important for testing models in a variety of classes of extragalactic objects.  However it is unavoidably the case that observational selection effects such as the flux limits and the positive redshift evolution of LFs in different wavebands makes determining the actual presence or absence, and the extent, and the form, of the intrinsic correlation between different waveband luminosities for a class of objects from flux-limited survey data complicated.  Figure \ref{firstfig} demonstrates that even intrinsically uncorrelated but flux-limited data can manifest observed luminosity correlations.  Our investigation is summarized below:

\begin{enumerate}

\item 
We first demonstrate analytically in \S \ref{analytic} the degree to which the observational selection effects and luminosity evolutions induce an artificial $L$--$L$ correlation and how this induced correlation is related to the parameters describing the LF, luminosity evolutions and co-moving density evolution. However, the bulk of the paper uses various simulated data sets to demonstrate the presence of induced correlations and the methods we use to determine true intrinsic correlations.

\item 
In \S \ref{sims} we describe three simulated populations of extragalactic sources with intrinsic properties (such as LF, luminosity evolutions in two different wavebands, and a common density evolutions) very similar to those deduced for quasars or AGNs, at radio (1.4 GHz) and optical (2500 \AA) bands; one with no correlation and two with two different degrees of correlations, quantified as $L'_{\rm rad} \propto (L'_{\rm opt})^{\alpha}$ with $\alpha$ = 0.0, 0.5 and 1.0, respectively.  We then select simulated ``observed'' sub-samples limited by two different flux limits in the two bands, again similar to observed quasar samples in the literature.

\item 
In \S \ref{partcor} we determined via the simulated data sets that considering full and partial correlations in bins of redshift is a useful method for determining presence or lack of, and at least qualitatively the relative degree, of intrinsic correlation between two waveband luminosities in a doubly flux-limited sample.    We also showed in \S \ref{bff} a technique to estimate the power-law form of the intrinsic correlation between luminosities.

\item 
In \S \ref{ep} we determined with the simulated data sets that non-parametric statistical techniques first proposed by \citet{EP92} and \citet{EP99} and extended to multiwavelength analyses in recent works such as \citet{QP1}, \citet{QP2}, \citet{BP2}, and \citet{QP3} can successfully recover the correct redshift evolutions of luminosities, redshift densities, and LFs of extragalactic populations catalogued in flux-limited surveys.  

\item 
In \S \ref{real} we demonstrated the techniques developed here for determining intrinsic $L$--$L$ correlations applied to two actual observed data sets.   Using binned partial correlation analysis we show that mid-infrared and optical luminosities show a stronger degree of intrinsic correlation than radio and optical luminosities. This is also manifested by larger index $\alpha$ of the intrinsic $L$--$L$ correlation for mid-infrared sample than the radio one.

The very high degree of correlation seen in this analysis between mid-infrared and optical luminosities in quasars lends support to the picture of tori being heated by primarily by accretion disks.  The significantly weaker correlation between radio and optical luminosities can be taken to support the notion that radio emission is affected by both the accretion disk size and the black hole spin, and maybe most importantly by the latter.  These results support an overall picture where black hole size determines accretion disk size and luminosity which then dominates the optical emission and becomes the primary driver of infrared emission via heating of the torus, while both black hole spin and size, and perhaps primarily spin, determine jet strength and therefore the radio luminosity.

\end{enumerate}

\acknowledgments

Funding for the SDSS and SDSS-II has been provided by the Alfred P. Sloan Foundation, the Participating Institutions, the National Science Foundation, the U.S. Department of Energy, the National Aeronautics and Space Administration, the Japanese Monbukagakusho, the Max Planck Society, and the Higher Education Funding Council for England. The SDSS Web Site is http://www.sdss.org/.  This publication makes use of data products from the Wide-field Infrared Survey Explorer, which is a joint project of the University of California, Los Angeles, and the Jet Propulsion Laboratory/California Institute of Technology, and {\it NEOWISE}, which is a project of the Jet Propulsion Laboratory/California Institute of Technology. {\it WISE} and {\it NEOWISE} are funded by the National Aeronautics and Space Administration.


\begin{thebibliography}{}

\bibitem[Antonucci(2012)]{Ski} Antonucci, R. 2012, AApTr, 27, 577
\bibitem[Becker et al.(1995)]{FIRST1} Becker, R., White, L., \& Helfand, D. 1995, \apj, 450, 559
\bibitem[Blandford \& Znajek(1977)]{BJ77} Blandford, R. \& Znajek, R. 1977,  \mnras, 179, 433
\bibitem[Blandford (1990)]{Blandford90} Blandford, R. 1990, ``Physical Processes in Active Galactic Nuclei.'' in Active Galactice Nuclei, ed. T. J.-L. Courvoisier \& M. Mayor (Berlin: Springer), 161
\bibitem[Broderick \& Fender(2011)]{BF11} Broderick, J.~W., \& Fender, R.~P. 2011, \mnras, 417, 184
\bibitem[Chanan (1983)]{Chanan83} Chanan, G. 1983, \apj, 275, 45
\bibitem[Efron \& Petrosian(1992)]{EP92} Efron, B. \& Petrosian, V. 1992, \apj, 399, 345
\bibitem[Efron \& Petrosian(1999)]{EP99} Efron, B. \& Petrosian, V. 1999, JASA, 94, 447', Mill Valley, CA: University Science Books 1989
\bibitem[Feigelson \& Berg(1983)]{FB83} Feigelson, E. \& Berg, C. 1983, \apj, 269, 400
\bibitem[Khembavi et al.(1986)]{Khembavi86} Khembavi, A., Feigelson, E., \& Singh, K. 1986, \mnras, 220, 51
\bibitem[Lawrence(1991)]{Lawrence91} Lawrence, A., 1991, \mnras, 252, 586
\bibitem[Lynden-Bell(1971)]{L-B71} Lynden-Bell, B. 1971, \mnras, 155, 95
\bibitem[Miller et al.(2010)]{Miller10} Miller, F., Vandome, A., \& McBrewster, J. 2010, ``Inverse Transform Sampling." (VDM Publishing)
\bibitem[Paige \& Carrera(2000)]{PC00} Feigelson, E. \& Berg, C. 2000, \mnras, 311, 433
\bibitem[Pavlidou et al.(2012)]{Pavlidou12} Pavlidou, V. et al., 2012, \apj, 751, 149
\bibitem[Petrosian(1992)]{P92} Petrosian, V. 1992, in Statistical Challenges in Modern Astronomy, ed. E.D. Feigelson \& G.H. Babu (New York:Springer), 173
\bibitem[Petrosian \& Singal(2015)]{CR1} Petrosian, V.,  \& Singal, J. 2015, ``On the Relation Between AGN Jet and Accretion Disk Emissions." in Extragalactic Jets From Every Angle, Proc. IAU S313, eds. F. Massaro, C. C. Cheung, E. Lopez, A. Siemiginowska (Cambridge: Cambridge University Press)
\bibitem[Rao \& Sievers (2007)]{RS07} Rao, S., \&  Sievers, G. 2007, ``A Rubist Partial Correlation Measure." {\it Journal of Nonparametric Statistics}, 5, 1
\bibitem[Richards et al.(2006)]{R06} Richards, G. et al. 2006, \aj, 131, 2766
\bibitem[Schneider et al.(2010)]{SDSSQ} Schneider, D., et al. 2010, \apj, 139, 2360
\bibitem[Sikora et al.(2007)]{Sikora07} Sikora, M., Stawarz, {\L}., \& Lasota, J.-P. 2007, \apj, 658, 815
\bibitem[Singal et al.(2011)]{QP1} Singal, J., Stawarz, {\L}., Lawrence, A., \&  Petrosian, V. 2011, ``On the Radio and Optical Luminosity Evolution of Quasars." \apj, 743, 104
\bibitem[Singal et al.(2012)]{BP1} Singal, Petrosian, V., \& Ajello, M. 2012, \apj, 753, 45
\bibitem[Singal et al.(2013)]{QP2} Singal, J., Stawarz, {\L}., Lawrence, A., \&  Petrosian, V. 2013, \apj, 764, 43
\bibitem[Singal et al.(2014)]{BP2} Singal, J., Ko, A., \&  Petrosian, V. 2014,  \apj, 786, 109
\bibitem[Singal (2015)]{BP3} Singal, J. 2015, \mnras, 115, 122
\bibitem[Singal et al.(2016)]{QP3} Singal, J., George, J., \&  Gerber, A. 2016, \apj, 831, 60
\bibitem[White et al.(1997)]{FIRST2} White, R., Becker, R., Helfand, D, \& Gregg., M. 1997, \apj, 475, 479


\end{thebibliography}
\end{document}